\documentclass[preprint]{aastex}

\usepackage{epsf}
\usepackage{emulateapj5}
\usepackage{onecolfloat}
\usepackage{apjfonts}
\usepackage{amsmath}

\def \xoff {\ifmmode x_{\rm off} \else $x_{\rm off}$ \fi}
\def \rhorms {\ifmmode \rho_{\rm rms} \else $\rho_{\rm rms}$ \fi}

\def \apj  {ApJ}
\def \apjs  {ApJS}

\def \mnras {MNRAS}
\def \etal {et~al.~}
\def \chisq  {\ifmmode  \chi^2   \else  $\chi^2$  \fi}  
\def \chisqr {\ifmmode \chi^2_{\rm r} \else $\chi^2_{\rm r}$ \fi}
\def \spose#1{\hbox  to 0pt{#1\hss}}  
\def \lta{\mathrel{\spose{\lower 3pt\hbox{$\sim$}}\raise  2.0pt\hbox{$<$}}}
\def \gta{\mathrel{\spose{\lower  3pt\hbox{$\sim$}}\raise 2.0pt\hbox{$>$}}}
 
\def \ha  {\ifmmode H\alpha \else H$\alpha $ \fi}

\def \kms {\ifmmode  \,\rm km\,s^{-1} \else $\,\rm km\,s^{-1}  $ \fi }
\def \kpc {\ifmmode  {\rm kpc}  \else ${\rm  kpc}$ \fi  }  
\def \Msun {\ifmmode M_{\odot} \else $M_{\odot}$ \fi} 
\def \hMsun {\ifmmode h^{-1}\,\rm M_{\odot} \else $h^{-1}\,\rm M_{\odot}$ \fi}
\def \hhMsun {\ifmmode h^{-2}\,\rm M_{\odot}\else $h^{-2}\,\rm M_{\odot}$ \fi}
\def \Lsun {\ifmmode L_{\odot} \else $L_{\odot}$ \fi} 
\def \hhLsun {\ifmmode h^{-2}\,\rm L_{\odot} \else $h^{-2}\,\rm L_{\odot}$ \fi}

\def \LCDM {\ifmmode \Lambda{\rm CDM} \else $\Lambda{\rm CDM}$ \fi}
\def \sig8 {\ifmmode \sigma_8 \else $\sigma_8$ \fi} 
\def \OmegaM {\ifmmode \Omega_{\rm M} \else $\Omega_{\rm M}$ \fi} 
\def \OmegaL {\ifmmode \Omega_{\rm \Lambda} \else $\Omega_{\rm \Lambda}$\fi} 
\def \Deltavir {\ifmmode \Delta_{\rm vir} \else $\Delta_{\rm vir}$ \fi}

\def \rs {\ifmmode r_{\rm s} \else $r_{\rm s}$ \fi} 
\def \rrm2 {\ifmmode r_{-2} \else $r_{-2}$ \fi} 
\def \ccm2 {\ifmmode c_{-2} \else$c_{-2}$ \fi} 
\def \cvir {\ifmmode c_{\rm vir} \else $c_{\rm vir}$ \fi} 
\def \cbar {\ifmmode \overline{c} \else $\overline{c}$ \fi}

\def \R200 {\ifmmode R_{200} \else $R_{200}$ \fi} 
\def \Rvir {\ifmmode R_{\rm vir} \else $R_{\rm vir}$ \fi}

\def \v200 {\ifmmode V_{200} \else $V_{200}$ \fi} 
\def \Vvir {\ifmmode V_{\rm  vir} \else  $V_{\rm vir}$  \fi} 
\def  \Vhalo  {\ifmmode V_{\rm halo} \else $V_{\rm halo}$ \fi}

\def \M200 {\ifmmode M_{200} \else $M_{200}$ \fi} 
\def \Mvir {\ifmmode M_{\rm  vir} \else $M_{\rm  vir}$ \fi}  
\def \Mshell  {\ifmmode M_{\rm shell} \else $M_{\rm shell}$ \fi}

\def \Nvir {\ifmmode N_{\rm  vir} \else $N_{\rm  vir}$ \fi}  

\def \Jvir {\ifmmode J_{\rm vir} \else $J_{\rm vir}$ \fi} 
\def \Jshell {\ifmmode J_{\rm shell} \else $J_{\rm shell}$ \fi}

\def \Evir {\ifmmode E_{\rm vir} \else $E_{\rm vir}$ \fi} 

\def \lam {\ifmmode \lambda  \else $\lambda$ \fi} 
\def \lamp {\ifmmode \lambda^{\prime} \else $\lambda^{\prime}$  \fi} 
\def \lampc {\ifmmode \lambda^{\prime}_{\rm c} \else
  $\lambda^{\prime}_{\rm c}$  \fi} 
\def \lambar {\ifmmode \bar{\lambda}  \else  $\bar{\lambda}$  \fi}  
\def  \lampbar  {\ifmmode \bar{\lambda^{\prime}} \else
  $\bar{\lambda^{\prime}}$\fi} 
\def \siglam {\ifmmode \sigma_{\lambda} \else $\sigma_{\lambda}$ \fi} 
\def \siglamp {\ifmmode                \sigma_{\lambda^{\prime}} \else
$\sigma_{\lambda^{\prime}}$\fi}

\def \Rd {\ifmmode R_{\rm d} \else $R_{\rm d}$ \fi} 
\def \Rs {\ifmmode R_{\rm s} \else $R_{\rm s}$ \fi}  
\def \Rd {\ifmmode R_{\rm d} \else $R_{\rm d}$ \fi}  
\def \Rcool  {\ifmmode R_{\rm  cool}  \else $R_{\rm cool}$ \fi} 
\def \RIII {\ifmmode  3.2\Rs \else $3.2\Rs$ \fi} 
\def \RII {\ifmmode 2.2\Rs \else $2.2\Rs$  \fi} 
\def \Reff {\ifmmode R_{\rm eff} \else $R_{\rm  eff}$ \fi} 
\def  \rb {\ifmmode r_{\rm b}  \else $r_{\rm b}$ \fi}

\def  \Sigmacrit   {\ifmmode  \Sigma_{\rm  crit}   
\else  $\Sigma_{\rm crit}$\fi} 
\def \Sig0 {\ifmmode \Sigma_{0} \else $\Sigma_{0}$ \fi}

\def \muI {\ifmmode \mu_{0,I} \else $\mu_{0,I}$ \fi}

\def \mgal {\ifmmode m_{\rm gal} \else $m_{\rm gal}$ \fi} 
\def \md {\ifmmode m_{\rm d} \else $m_{\rm d}$ \fi} 
\def \ms {\ifmmode m_{\rm   s}   \else   $m_{\rm   s}$   \fi}   
\def   \mdbar   {\ifmmode {\overline{m}}_{\rm d} \else
  ${\overline{m}}_{\rm d}$ \fi} 
\def \msbar {\ifmmode  \bar{m}_{\rm  s}  \else  $\bar{m}_{\rm s}$
  \fi}  
\def  \Md {\ifmmode M_{\rm d}  \else $M_{\rm d}$ \fi} 
\def  \Ms {\ifmmode M_{\rm s} \else $M_{\rm  s}$ \fi} 
\def \Mb {\ifmmode  M_{\rm b} \else $M_{\rm b}$ \fi} 
\def \Mstar {\ifmmode  M_{\rm star} \else $M_{\rm star}$ \fi}
\def \Mdisc {\ifmmode M_{\rm disc} \else $M_{\rm disc}$ \fi}

\def \Jd {\ifmmode J_{\rm d} \else $J_{\rm d}$ \fi} 
\def \Jb {\ifmmode J_{\rm b} \else $J_{\rm b}$ \fi}  
\def \fb {\ifmmode  f_{\rm b} \else $f_{\rm b}$ \fi}

\def  \jd  {\ifmmode j_{\rm  d}  \else  $j_{\rm  d}$ \fi}  
\def  \jdmd {\ifmmode \frac{j_{\rm  d}}{m_{\rm d}} \else
  $\frac{j_{\rm d}}{m_{\rm d}}$ \fi} 
\def \fj {\ifmmode f_{\rm j} \else $f_{\rm j}$ \fi} 
\def \ft {\ifmmode f_{\rm t}  \else $f_{\rm t}$ \fi} 
\def  \fM {\ifmmode f_{\rm M} \else $f_{\rm M}$ \fi}

\def  \Vd {\ifmmode  V_{\rm  d}  \else $V_{\rm  d}$  \fi} 
\def  \Vcool {\ifmmode V_{\rm cool} \else $V_{\rm cool}$ \fi} 
\def \Vcirc {\ifmmode V_{\rm circ}  \else $V_{\rm circ}$  \fi} 
\def \VIII  {\ifmmode V_{3.2} \else $V_{3.2}$ \fi} 
\def  \VII {\ifmmode V_{2.2} \else $V_{2.2}$ \fi}
\def \Vobs {\ifmmode V_{\rm obs}  \else $V_{\rm obs}$ \fi} 
\def \Vdisc {\ifmmode V_{\rm disc} \else  $V_{\rm disc}$ \fi} 
\def \Vmax {\ifmmode V_{\rm  max} \else  $V_{\rm max}$  \fi} 
\def  \Vmaxobs{\ifmmode V_{\rm max}^{\rm obs}\else  $V_{\rm max}^{\rm
    obs}$\fi}  
\def \Vtot {\ifmmode V_{\rm tot} \else $V_{\rm tot}$  \fi} 
\def \Vrot {\ifmmode V_{\rm rot} \else  $V_{\rm rot}$  \fi} 
\def  \Vflat {\ifmmode  V_{\rm  flat} \else $V_{\rm flat}$ \fi}

\def \Ups {\ifmmode \Upsilon  \else $\Upsilon$ \fi} 
\def \YB {\ifmmode \Upsilon_B \else $\Upsilon_B$ \fi} 
\def \YI {\ifmmode  \Upsilon_I  \else $\Upsilon_I$ \fi} 
\def \DeltaIMF {\ifmmode \Delta_{\rm IMF} \else $\Delta_{\rm IMF}$ \fi}

\def\LCDM{$\Lambda$CDM }

\def\c200{$c_{200}$}

\begin{document}
\submitted{The Astrophysical Journal, submitted}
\vspace{1mm}
\slugcomment{{\em The Astrophysical Journal, submitted}}

\shortauthors{MACCI\`O ET AL.}
\twocolumn[
\lefthead{Central mass and luminosity of Milky Way satellites}
\righthead{Macci\`o et al.}

\title{Central mass and luminosity of Milky Way satellites in the \LCDM model}

\author{Andrea V. Macci\`o\altaffilmark{1}, Xi Kang\altaffilmark{1},
Ben Moore\altaffilmark{2} 
}

\begin{abstract}

  It has been pointed out that the Galactic satellites all have
  a common  mass around $10^{7} M_{\odot}$ within  300 pc ($M_{0.3}$),
  while they span  almost four order of magnitudes  in luminosity 
  (Mateo \etal 1993, Strigari et al. 2008).  It
  is  argued  that this  may  reflect  a  specific scale  for  galaxy
  formation  or  a scale  for  dark  matter  clustering. Here  we  use
  numerical simulations coupled with  a semi-analytic model for galaxy
  formation, to  predict the central  mass and luminosity  of galactic
  satellites in the $\Lambda$CDM model.  We show that this common mass
  scale can be explained within the Cold  Dark Matter scenario
  when the  physics of  galaxy formation is  taken into  account.  The
  narrow  range of  $M_{0.3}$ comes  from the  narrow  distribution of
  circular velocities  at time of  accretion (peaking around  20 km/s)
  for  satellites able  to form  stars and the not tight correlation
  between halo concentration and circular velocity. 
  The  wide range  of satellite  luminosities is due to a 
combination of the  mass at time of accretion and 
the broad distribution of accretion redshifts for a  given mass.
  This causes  the satellites baryonic
  content   to  be   suppressed  by   photo-ionization   to  very different
  extents.  Our  results  favor  the  argument  that  the  common  mass
  $M_{0.3}$  reflects a  specific scale  (circular velocity  $\sim 20$
  km/s ) for star formation.

\end{abstract}

\keywords{cosmology: theory --- galaxies: dwarf --- hydrodynamics --- methods: numerical}
]

\altaffiltext{1}{Max-Planck-Institut f\"ur Astronomie, K\"onigstuhl 17, 69117
  Heidelberg, Germany; maccio@mpia.de}
\altaffiltext{2}{Institute for Theoretical Physics, University of Z\"urich,
Winterthurerstrasse 190, CH-8057 Z\"urich, Switzerland.}

%%%%%%%%%%%%%%%%%%%%%%%%%%%%%%%%%%%%%%%%%%%%%%%%%%%%%%%%%%%%%%%%%%%%%%
%% SECTION 1: INTRODUCTION
%%%%%%%%%%%%%%%%%%%%%%%%%%%%%%%%%%%%%%%%%%%%%%%%%%%%%%%%%%%%%%%%%%%%%%
\section{Introduction}
\label{sec:intro}

Galaxies are thought  to form out of gas which  cools and collapses to
the center of  dark matter haloes (White \& Rees 1978). A correlation
between  stellar  mass and  host  halo  mass  is thus  expected.  Observational data, such as
from the SDSS, have shown  a  good correlation  between the
stellar mass  of central galaxies and the halo mass of galaxy groups (e.g.
Yang et  al. 2007), but  such a correlation  is not mantained in satellite
galaxies (e.g., Gao  et al. 2004, Conroy et al. 2007).  It is also not
clear if  such a relation persists to  all mass scales.  For example, in
galaxy  clusters, the  stellar mass  of the  central galaxy  is  not a
strong  function  of the  halo  mass,  because  gas cooling  and  star
formation are regulated by the physical processes operating in clusters,
such as AGN feedback, which gives rise to
a steep luminosity function at the bright end (e.g., Kang et al. 2006).

Mateo et al. (1993) and more recently, Strigari  et  al. 
(2008,  hereafter  S08), claimed that all the
satellites with luminosity between  $10^3$ and $10^7 L_{\odot}$ to have
a  common mass  of  $\sim 10^7  M_{\odot}$  within a  radius of  300pc
($M_{0.3}$), and to  be dark matter dominated within  this region.   
This narrow range for central  masses compared to
the wide  range of  luminosities has raised  the issue if  this common
mass may reflect a specific scale for galaxy formation or a scale for
dark matter clustering.

In this letter,  we use a series of  high-resolution simulations of MW
type haloes combined with  a semi-analytical model of galaxy formation
to show how  this flat relation between $M_{0.3}$  and luminosity arises
naturally within our standard cosmological framework combined with simple 
modelling of the galaxy formation process.
\\
\\

%%%%%%%%%%%%%%%%%%%%%%%%%%%%%%%%%%%%%%%%%%%%%%%%%%%%%%%%%%%%%%%%%%%%%%
%% SECTION 2: N-BODY SIMULATIONS
%%%%%%%%%%%%%%%%%%%%%%%%%%%%%%%%%%%%%%%%%%%%%%%%%%%%%%%%%%%%%%%%%%%%%%
\section{Simulations}
\label{sec:sims}

Nbody simulations  have been carried out using {\sc  pkdgrav}, a treecode
written  by  Joachim  Stadel  and  Thomas Quinn  (Stadel  2001).   The
cosmological  parameters are set  in agreement  with the  WMAP mission
first   year   results   (WMAP1: Spergel   \etal   2003)   as follows:
$\Omega_{\Lambda}$=0.732, $\Omega_m$=0.268, $\Omega_b$=0.044, $h=0.71$,
$n=1.0$ and $\sigma_8=0.9$.   
We have selected  three candidate haloes  with a
mass similar to  the mass of our Galaxy ($M  \sim 10^{12} \Msun$) from
an existing  low resolution dark matter  simulation (300$^3$ particles
within 90  Mpc) and re-simulated  them at higher resolution  using the
volume  renormalization  technique (Katz \& White 1993).  Our high
resolution haloes all have a quiet  merging history with no  major merger
after $z=2$, thus are likely to host a disk galaxy at the present time
(as confirmed by our semi-analytic models, see section \ref{sec:sam})
The high  resolution run  is 12$^3$ times  better resolved than  the low
resolution one: the dark matter particle mass is $m_{d} = 4.16 \times 10^5
h^{-1}  \Msun$,   where  each  dark  matter  particle   has  a  spline
gravitation  (comoving) softening  of 355  $h^{-1}$ pc.  Properties of
single halos are listed in Table \ref{table:gal}.

For  the  purpose  of  constructing  accurate merger  trees  for  each
simulated galaxy  we analyzed 53 outputs between $z=20$
and  $z=0$.  For  each snapshot  we looked  for all  virialized haloes
within  the  high  resolution  region using  a  spherical  overdensity
algorithm (see Macci\`o  \etal  2007  for more  details  on our  halo
finding procedure).
We included  in the halo catalogue all  haloes with more than
100 particles  ($\ge 4\times  10^{7} h^{-1} \Msun$).  For the  merger tree
construction we started marking all the particles within the 1.5 times
the virial radius of a given  galaxy at $z=0$ and we tracked them back
to the previous output time. We then make a list of all haloes that at
earlier output time containing  marked particles, recording the number
of marked particles. In addition we record the
number of  particles that are not  in any halo in  the previous output
time and we consider them as {\it smoothly} accreted.  We used the two
criteria suggested in  Wechsler \etal (2002) for halo  1 at one output
time to be labeled a ``progenitor'' of halo 2 at the subsequent output
time.  In our language halo 2 will then be labeled as a ``descendant''
of halo 1 if i) more than 50\% of the particles in halo 1 end up
in halo  2 or  if ii) more than 75\% of halo 1 particles that end
up in any halo at time step 2 do end up in halo 2 (this second 
criterion is mainly relevant during major mergers).
A halo can have  only one  descendant but there  is no
limit  to the  number of  progenitors.  On average  there are  ~20,000
progenitors for each high resolution DM halo.

%%%%%%%%%%%%%%%%%%%%%%%%%%%%%%%%%%%%%%%%%%%%%%%%%%%%%%%%%%%%%%%%%%%%%%%%
%% SECTION 3: SAMs
%%%%%%%%%%%%%%%%%%%%%%%%%%%%%%%%%%%%%%%%%%%%%%%%%%%%%%%%%%%%%%%%%%%%%%%%
\section{Semi Analytical Models}
\label{sec:sam}

We   calculate   the   luminosity   of  the   satellites   using   the
semi-analytical model  for galaxy formation of Kang  \etal (2005), and
we refer  the reader  to this paper  for more details.  Basically we
follow the formation  history of dark matter  haloes and graft the
physical models for galaxy  formation onto the merger trees.  Physical
processes  governing  galaxy   formation  include:  gas  cooling  from
radiative  hot  gas,  star  formation  and  supernova  feedback.   The
original  model  has been  recently  updated  (Kang  2008) to  include
photo-ionization  to  suppress baryon  accretion  in  low mass  haloes
(Kravtsov et al.  2004), along  with a new fitting formula to describe
dynamical friction  time-scales for  satellite galaxies (Jiang  et al.
2008).  Kang  (2008)  has  shown  that this  model  is  successful in
reproducing  the Milky  Way  satellite luminosity  function and  other
properties (Macci\`o \etal in preparation).

\begin{table}
\caption{Galaxies Parameters}
\begin{tabular}{lccccc}
\hline  Halo &  Mass  & Npart  & $R_{vir}$ & MaxVcirc  \\
       &  ($10^{12}\hMsun)$      &        &  (kpc/h)  & (km/s) \\
\hline 
G0   & 0.88  & 2115385   & 197 & 179.6  \\
G1   & 1.22  & 2931295   & 219 & 187.6  \\
G2   & 1.30  & 3123511   & 250 & 203.1  \\
\hline 
\label{table:gal}
\end{tabular}
\end{table}

There  are  two  important   factors  which  regulate  star  formation
efficiency in low  mass dark matter haloes. Firstly,  the cooling rate
for hot gas in haloes with a virial temperature $T_{vir} <  10^{4}K$ 
is quite low due to the inefficiency  of H$_{2}$ cooling, 
thus we shut off gas cooling in such haloes. 
This  effect is responsible for the  absence of visible
satellites   with  $\Vcirc   \lta  16.6$   km/s  as   shown   in  Figure
\ref{fig:vcirc2}.   Secondly, hot  gas  accretion is  suppressed by  the
cosmic  photo-ionization background  in low  mass haloes;  the typical
filtering-mass, at which haloes  lose half of their baryonic content, 
has a strong redshift dependence 
and it increases from $10^{7}M_{\odot}$ at $z \sim
10$ to  $3\times 10^{10}M_{\odot}$ at  $z=0$ (Kravtsov et  al.  2004).
This  implies  that  a  satellite  galaxy with a halo  mass  around
$10^{9}M_{\odot}$  at infall,  has  its baryon  content suppressed  to
different extents depending on its  accretion redshift.  We will see in
Section \ref{sec:res}  that this  gives  rise to  the wide  luminosity
range  for satellites, despite  the narrow  range spanned  by circular
velocities at the accretion time. We set the reionization redshift to be 
$z_r=7$  but as shown by 
Kravtsov \etal (2004) the results of photo-ionization are 
almost insensitive to the actual value of $z_r$.

%%%%%%%%%%%%%%%%%%%%%%%%%%%%%%%%%%%%%%%%%%%%%%%%%%%%%%%%%%%%%%%%%%%%%%%%
%% SECTION 4: Results SAMs
%%%%%%%%%%%%%%%%%%%%%%%%%%%%%%%%%%%%%%%%%%%%%%%%%%%%%%%%%%%%%%%%%%%%%%%%
\section{Results}
\label{sec:res}

The relation between the central  mass of a satellite and its circular
velocity  can be  predicted on  a  theoretical basis,  using the  well
studied relation  between mass and  concentration.  N-body simulations
have  shown that  the spherically  averaged density  profiles  of dark
matter  haloes can  be well  described  by a  two parameter  analytic
profile (Navarro, Frenk \& White 1997, NFW hereafter):
\begin{equation}
\frac{\rho(r)}{\rho_{\rm crit}} = \frac{\delta_{\rm c}}{(r/\rs)(1+r/\rs)^2},
\label{eq:nfw}
\end{equation}
where  $\rho_{\rm crit}$  is  the critical  density  of the  universe,
$\delta_{\rm c}$  is the characteristic  overdensity of the  halo, and
$\rs$ is  the radius where the  logarithmic slope of  the halo density
profile  ${\rm d}\ln\rho/{\rm  d}\ln r  =  -2$.   A more  useful
parametrization  is  in  terms  of  the  virial  mass,  $\Mvir$,  and
concentration  parameter,  $c\equiv\Rvir/\rs$.   The virial  mass  and
radius are related by $\Mvir = \Delta_{\rm vir} \rho_{\rm crit} (4 \pi
/ 3) \Rvir^3$, where $\Delta_{\rm vir}$ is the density contrast of the
halo.
Then the mass within 300 pc is simply defined by:
\begin{equation}
  M_{0.3} = \int_0^{r_{0.3}} 2 \pi t^2 \rho(t) dt = 4 \pi \rho_{\rm crit} \delta_{\rm c} \rs^3 (\log(1+x)- \frac {x}
  {1+x})
\label{eq:m03}
\end{equation}
where $x \equiv  r_{0.3}/\rs$ and the value of  $M_{0.3}$ only depends
on  the two  parameters defining  the density  profile: $\rs$  (or the
concentration $c$)  and $ \delta_{\rm  c}$.  Several models  have been
proposed to link mass and concentration of dark matter haloes (Bullock
\etal 2001,  Eke \etal 2001,  Macci\`o \etal 2007),  recently Macci\`o
\etal 2008 have  proposed a new toy model (based  on a modification of
original  model of Bullock  \etal 2001)  to predict  the concentration
mass  relation in dark  matter haloes.   Using this  model it  is then
possible  to predict  the  value of  $M_{0.3}$  as a  function of  the
circular velocity of the halo.

\begin{figure}
\plotone{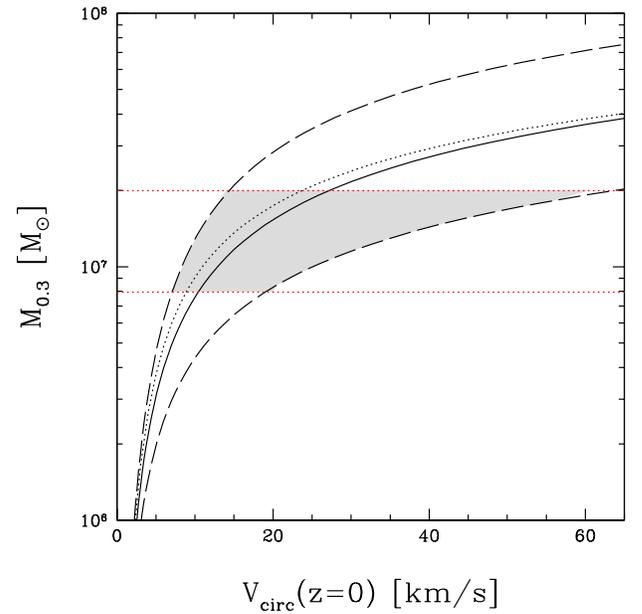}
\caption{ Theoretical prediction of  the mass within 300 pc
  versus circular velocity  using the toy model of  Macci\`o \etal 2008
  for  the mass  concentration  relation: solid line and dashed lines
  represent the median and two $\sigma$ scatter respectively for the 
  WMAP5 model (the dotted line shows the median for the WMAP1 model).
  The dotted (red) horizontal lines  show  the mass  range  observed Milky  Way
  satellites (S08).   The shaded region  gives a visual  impression of
  circular velocities for  which the values of $M_{0.3}$ consistent
  with observations are expected.}
\label{fig:vcirc1}
\end{figure}

Figure \ref{fig:vcirc1} shows the values for the mass within 300 pc in
the WMAP5 cosmology (Komatsu \etal 2008) as a function of the circular
velocity of the dark matter halo.  Dashed lines represent the one $\sigma$
scatter as  expected from  Nbody simulations (here  we use  a constant
value of 0.1  dex Macci\`o \etal 2008).  The  dotted horizontal lines
show the mass range observed for Milky Way satellites (S08). The range
of  circular  velocities  that  is  in  agreement  with  observational
data  (shaded gray  region in  the  figure) is  fairly large  and
covers almost the entire range of plausible values for $\Vcirc$.

%-----------------------------------------------------------------------------------------------

%-----------------------------------------------------------------------------------------------

There are two possible issues with this kind of approach: firstly, we are
extrapolating to  lower masses a relation ($c/M$) that  has been tested
on N-body simulation only down to  a mass of $\sim 10^{10} \Msun$,
well above  the expected masses  for Milky Way  satellites. Secondly,
satellites we see today can  be a biased  sample of the  overall dark
matter halo  population since they are the surviving population 
which possibly form prior to reionisation (Moore et al. 2006).   
For these reasons  a full numerical
inspection of origin of the narrow range for $M_{0.3}$ is needed.

%-----------------------------------------------------------------------------------------------
\begin{figure}
\plotone{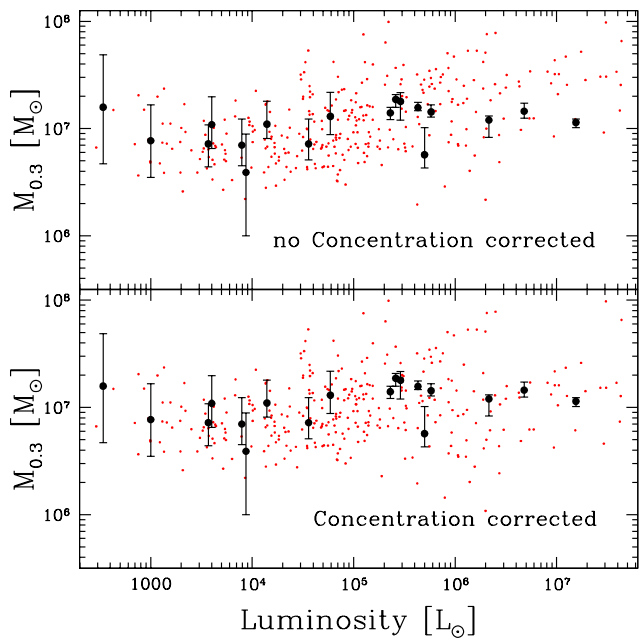}
\caption{Mass  within 300 pc versus  luminosity.  Red dots
  show results from our numerical  model, black points with error bars
  are the  observational results from S08. Upper  panel: no correction
  for  the  concentration  related  density  evolution.  Lower  panel:
  correction included (see text).}
\label{fig:str1}
\end{figure}
%-----------------------------------------------------------------------------------------------

In  order  to compare  numerical  results  to  observations, for  each
simulated satellite  we need  both its luminosity  $L$ and  its inner
mass $M_{0.3}$.   The first quantity is  a direct outcome  of our 
semi-analytical model, whilst  to compute the second quatity we proceeded in the
following way: at  time of accretion  of each  satellite we compute
the  density profile directly  from its  particle distribution  in the
N-body simulation.   The resulting  numerical density profile  is then
fitted  with  an  NFW   profile  (Eq.~\ref{eq:nfw});  during  the  fitting
procedure we treat both $\rs$ and $\delta_c$ as free parameters. Their
values,  and associated  uncertainties,  are obtained  via a  $\chi^2$
minimization procedure (see Macci\`o \etal 2008 for more details).  We
are only interested in dark matter haloes that host a galaxy according
to our semi-analytic model; given  that gas cooling is allowed only in
haloes with  $M\gta 10^8  \hMsun$ (i.e. $T_{vir}>10^4$K)  this implies
that, on average,  we have more than 1,000 particles  per halo at time
of accretion, which is sufficient to obtain a robust estimation of the
density   profile  parameters  (Macci\`o   \etal  2007).    Under  the
assumption that the density profile within 300 pc does not evolve from
the time  of accretion to  $z=0$ we can  compute, for each satellite,
the present  value of $M_{0.3}$  using equation \ref{eq:m03}.

The  upper panel  in Figure  \ref{fig:str1} shows  the results  for the
relation between the mass within 300  pc and luminosity as obtained in our
numerical model (red dots)  versus the observational results (black
dots  with  error bars).  Here  we  plot  results only  for  simulated
satellites  that  satisfy  the  detection  threshold of  the  SDSS  as
determined by Koposov \etal 2007. This means that the satellite luminosity
and    distance    have   to    satisfy    the   following    relation
$\log(R/{\rm kpc})<1.04-0.228~M_v$.    The   mean   and   the   scatter   of
observational data  are both well reproduced by  our numerical results
up  to  a  luminosity  of   {$L=2\times  10^6 \Lsun $},  after  this  point
simulations seem to suggest  an increase with luminosity of $M_{0.3}$,
which is  not present in the  data (even if only three satellite galaxies
have a luminosity greater than $10^6 \Lsun$).

These results  are obtained under  the assumption of no  evolution for
the parameters  defining the  density profile ($\rs {\rm and} \delta_c$).
This  assumption is motivated  by the  detailed  numerical study
carried out  by Kazantzidis \etal  2004 (K04, hereafter, see  also the
recent results by and Pe{\~n}arrubia \etal 2008).  
K04 have shown that the
inner density profile is extremely robust and that it is unmodified by
tidal forces  even after tidal  stripping removes a large  fraction of
the  initial  mass.    They  have  also  shown  that   the  degree  of
modification (if  any) of the  density profile depends on  the initial
(i.e.  before infall) concentration of the satellite dark matter halo.
While highly concentrated  haloes (with $c \gta 15$)  are able to keep
the  profile  unchanged  even  after  several  orbital  periods,  less
concentrated  haloes ($c  \approx  9$) slightly  modify their  profile
mainly   by  reducing  the   overall  normalization   ($\delta_c$)  by
approximately a factor 2 (see also Mayer \etal 2006).

In  order to  take  into  account this  expected  modification of  the
density profile  in low concentration  haloes, we manually  reduce the
$\delta_c$ parameter  by a factor  2 in all haloes  with concentration
less than 10  at the moment of infalling, In  doing this correction we
used the value of the concetration extrapolated to z=0, in other words
we  multiplied   the  value  of  the  concetration   at  $z_{acc}$  by
$(1+z_{acc})$ in order to take  into account the redshift evolution of
$\Rvir$. Results  are shown in  lower panel of  Figure \ref{fig:str1}.
As  expected  this  modification  mainly applies  to  high  luminosity
haloes, since they  were the most massive ones at  time of infall, and
thus likely to be less  concentrated (the average concentration of our
haloes  is $16.3 \pm  7.1$, and  74\% have  $c>10$).  When  a possible
modification of  the density profile  for low concentration  haloes is
taken into account the up turn in the numerical $M_{0.3}$/$L$ relation
at high luminosities almost vanishes  and numerical results are now in
better agreement with the observational data. There are still some haloes
with $M_{0.3}>3 \times 10^7$ around $L=10^6$, these haloes formed at high 
redshift, and thus happen, by chance, to have a high $c$ (i.e. they are not affected 
by our correction) and a large $L$; besides of that they do not present any
other peculiar behavior.

The presence of  a baryonic component inside the  dark matter halo can
by itself modify the density profile of the halo, due to the adiabatic
compression  process   (e.g.  Blumenthal  \etal   1986). 
In our  case we expect this effect to  be negligible given that
the baryon fraction of our haloes is much lower then the universal one
and satellites have been observed  to be dark matter dominated even in
their central regions (S08).

%-----------------------------------------------------------------------------------------------
\begin{figure}
\plotone{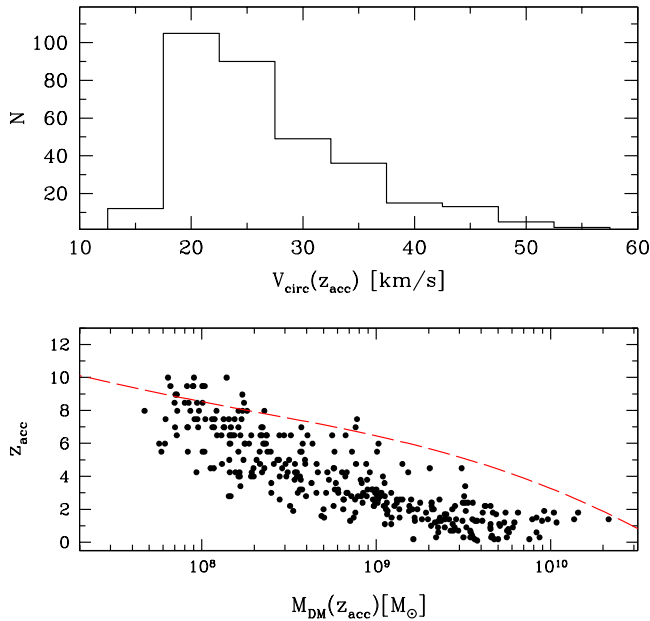}
\caption{  Upper   panel:  distribution  of   the  circular
  velocities  at the  time of  accretion for  visible  satellites. Lower
  panel:  accretion   redshift  vs   satellite  halo   mass  at
  accretion time. The  (red) dashed line shows  the redshift evolution 
  of the filtering mass for UV photoionization.}
\label{fig:vcirc2}
\end{figure}
%-----------------------------------------------------------------------------------------------

Now we turn to understand the origin of the relation between $M_{0.3}$
and $L$  found by  S08. In upper  panel of Figure  \ref{fig:vcirc2} we
show  the distribution  of $V_{circ}$  at  the time  of accretion  for
visible satellites  (same sample used in  Figure \ref{fig:str1}).  The
distribution  peaks around $\Vcirc=20$ km/s  and then declines  sharply 
towards  higher values  of  the circular  velocity.   As discussed  in
Section \ref{sec:sam}, the  sharp cutoff  below $\Vcirc \sim  20$ km/s
comes from the  shut off of cooling in  haloes with virial temperature
below  $10^4$K.   Combining   this  narrow  distribution  of  circular
velocity  (between  $20-40$ km/s)  with  the theoretical  expectations
shown in Figure \ref{fig:vcirc1}, and assuming that $M_{0.3}$ does not
evolve  after accretion,  it  is  not surprising  that  all Milky  Way
satellites (observed  and simulated) have  a inner mass within  300 pc
always around  $10^7 \Msun$. It is  then interesting to  ask why these
satellites span  a wide  range of luminosity.   In the lower  panel of
Figure \ref{fig:vcirc2}, we show the dark matter mass of satellites at
their time of accretion, and the dashed line shows the evolution of 
the filter mass,
defined as the mass of a halo in which half of its baryons have been lost
due to photo-ionization.  
We can  see that most satellites have  mass lower than
the filter  mass at accretion, and  at a given accretion  mass, there is
wide range of accretion redshifts,  which give rise to different amounts
of baryon suppression.
The spread in Luminosity orginates then from the range of halo masses
at time of accretion combined with the large scatter (at a given mass)
in the accretion time.

Recently Okamoto, Gao \&  Theuns (2008) have
shown that the actual values of the filter mass might be smaller 
than  what suggested by Kravtsov \etal (2004). The effect of a lower
filtering mass will be to increase the satellite luminosity 
for a given $\Vcirc$; this will push all points in 
Figure \ref{fig:str1} towards the right; but will not alter the flat 
relation between $10^{4}L_{\odot}$ and $10^7L_{\odot}$.

Whilst this paper was ready for submission a similar study was posted
on the  arXiv (Li \etal 2008). These authors used  a similar approach
(an Nbody  simulation combined with  a semi-analytic model  for galaxy
formation) to  study the relation between central  mass and luminosity
in  (satellite) dwarf galaxies.  The main  difference with  respect to
this  work  is  that  Li  \etal  (2008) computed  the  inner  mass  of
satellites  directly  from the  Nbody  simulation.  Given the  spatial
resolution of  their simulation (0.18 kpc) they  presented results for
$M_{0.6}$ (mass  within 600 kpc) which  turned out to  be in excellent
agreement  with observational  data  of Strigari  \etal (2007).   This
study   is  complementary   to   ours.

%%%%%%%%%%%%%%%%%%%%%%%%%%%%%%%%%%%%%%%%%%%%%%%%%%%%%%%%%%%%%%%%%%%%%%
%% SECTION 5: SUMMARY
%%%%%%%%%%%%%%%%%%%%%%%%%%%%%%%%%%%%%%%%%%%%%%%%%%%%%%%%%%%%%%%%%%%%%%
\section{Summary and Conclusions}
\label{sec:concl}

The observational evidence that all Milky Way satellites have a
common mass of about $10^7 \Msun$ within their central 300 parsecs has
raised issues  about the  possible existence  of a  new  scale in
galaxy formation or a characteristic  scale for the clustering of dark
matter.   In   this  Letter,   by  using  high   resolution  numerical
simulations combined with a  semi analytic model for galaxy formation,
we show that this common mass scale can be easily explained within the
current $(\Lambda$)CDM model for structure formation.

The  observational   data  on   the  $L$/$M_{0.3}$  relation   can  be
successfully reproduced in numerical simulations, up to a luminosity of
$10^6 \Lsun$ under the assumption that the 
parameters  describing the density profile  of satellite galaxies do 
not evolve significantly
after these galaxies have been accreted into the main halo.  When
a  plausible   (small)  modification   of  such  parameters,   for  low
concentration  haloes,  is  taken  into account,  the  agreement  with
observational data extends to the whole luminosity range. 

According to  our numerical modeling this common  mass scale, $M_{0.3}
\sim  10^7  \Msun$,  originates  from  the narrow  range  of  circular
velocities (20-40 km/s)  spanned by visible satellites at  the time of
accretion (with  a lower limit of  17 km/s set by  the inefficiency of
H$_{2}$ cooling in halo with virial temperature below $10^{4}$ K).  On
the other  hand the  wide range of  luminosities comes from  
the range of halo masses at time of accretion combined with 
the large scatter  in the  accretion  time for  an  halo with  
a given  mass, which has  a  strong  impact  on  the   ability  of  photo
reionization of  reducing the baryon  content of satellites,  and thus
determining  their luminosity.   

Our  results show that  the observed  flat relation  between satellite
inner  mass and  luminosity can  be  easily explained  within the  CDM
framework, without modifying the nature of dark matter particles. 

\acknowledgements

The  authors  thank  L.  Strigari  for  making  his  results  publicly
avaliable.   Frank van  den Bosch  and Sergey  Koposov  are gratefully
acknowledged  for  interesting  discussions  and useful  comments.  An
anonymous referee  is also thanked for his/her  comments that improved
the presentation of this  paper.  Numerical simulations were performed
on on the  PIA cluster of the Max-Planck-Institut  f\"ur Astronomie at
the Rechenzentrum in Garching.

%%%%%%%%%%%%%%%%%%%%%%%%%%%%%%%%%%%%%%%%%%%%%%%%%%%%%%%%%%%%%%%%%%%%%%
%%  REFERENCES
%%%%%%%%%%%%%%%%%%%%%%%%%%%%%%%%%%%%%%%%%%%%%%%%%%%%%%%%%%%%%%%%%%%%%% 


\begin{thebibliography}{}



\bibitem[Blumenthal   et  al.(1986)]{1986ApJ...301...27B}  Blumenthal,
G.~R., Faber, S.~M.,  Flores, R., \& Primack, J.~R.\  1986, \apj, 301,
27


\bibitem[Bullock  et al.(2001)]{2001MNRAS.321..559B}  Bullock, J.~S.,
  Kolatt,  T.~S.,  Sigad,  Y.,  Somerville,  R.~S.,  Kravtsov,  A.~V.,
  Klypin, A.~A., Primack, J.~R., \&  Dekel, A.\ 2001, MNRAS, 321, 559
  (B01)

\bibitem[Conroy et al. 2007]{C07}
Conroy, C., Wechsler, R.H., \& Kravtsov, A.V., 2007, ApJ, 668, 826

\bibitem[Eke et al.(2001)]{2001ApJ...554..114E} Eke, V.~R., Navarro, J.~F., 
\& Steinmetz, M.\ 2001, \apj, 554, 114 

\bibitem[Gao et al. 2004]{Gao04}
Gao, L., De Lucia, G., White, S. D. M., \& Jenkins, A., 2004, MNRAS, 352, 1

\bibitem[Jiang et al. 2008]{J08}
Jiang, C.Y., Jing, Y.P., Faltenbacher, A., Lin, W.P., \& Li, C., 2008, ApJ, 675, 1095

\bibitem[Kang et al. 2005]{K05}
Kang, X., Jing, Y.P., Mo, H.J., B\"orner, G., 2005, ApJ, 631, 21

\bibitem[Kang et al. 2006]{K06}
Kang, X., Jing, Y.P., Silk, J., 2006, ApJ, 648, 820

\bibitem[Kang 2008]{K08}
Kang, X., 2008, 
Proceedings of IAU 254 "The Galaxy Disk in Cosmological Context", arXiv:0806.3279

\bibitem[Katz \& White(1993)]{1993ApJ...412..455K} 
Katz, N., \& White, S.~D.~M.\ 1993, \apj, 412, 455 

\bibitem[Kazantzidis et al.(2004)]{2004ApJ...608..663K} Kazantzidis, S., 
Mayer, L., Mastropietro, C., Diemand, J., Stadel, J., 
\& Moore, B.\ 2004, \apj, 608, 663 

\bibitem[Komatsu et al.(2008)]{2008arXiv0803.0547K} Komatsu, E., et al.\ 
2008, arXiv:0803.0547 

\bibitem[Koposov et al.(2008)]{2007arXiv0706.2687K} Koposov, S., et al.\ 
  2008, ApJ, 686, 279

\bibitem[Kravtsov et al. 2004]{K04}
Kravtsov, A.V., Gnedin, O.Y., \& Klypin, A.A., 2004, ApJ, 609, 482

\bibitem[Li et al.(2008)]{2008arXiv0810.1297L} Li, Y.-S., Helmi, A., De 
Lucia, G., \& Stoehr, F.\ 2008, arXiv:0810.1297 

\bibitem[Macci{\`o} et al.(2007)]{2007MNRAS.378...55M} Macci{\`o}, A.~V., 
Dutton, A.~A., van den Bosch, F.~C., Moore, B., Potter, D., \& Stadel, J.\ 
2007, MNRAS, 378, 55 

\bibitem[Macci{\`o} et al.(2008)]{2008MNRAS} Macci{\`o}, A.~V., 
Dutton, A.~A., van den Bosch, F.~C., 2008, MNRAS in press, arXiv:0805.1926 

\bibitem[Mayer et al.(2006)]{2006MNRAS.369.1021M} Mayer, L., Mastropietro, 
C., Wadsley, J., Stadel, J., \& Moore, B.\ 2006, \mnras, 369, 1021 

\bibitem[Mateo et al.(1993)]{1993AJ....105..510M} Mateo, M., Olszewski, 
E.~W., Pryor, C., Welch, D.~L., \& Fischer, P.\ 1993, \aj, 105, 510 

\bibitem[Moore et al.(2006)] {2006mo} Moore, B., Diemand, J., Madau, P., Zemp, M. \& Stadel, J. 2006, MNRAS, 368, 563

\bibitem[Navarro  et  al.(1997)]{1997ApJ...490..493N} Navarro,  J.~F.,
Frenk, C.~S., \& White, S.~D.~M.\ 1997, ApJ, 490, 493

\bibitem[Okamoto et al. 2008]{Oka08}
Okamoto, T., Gao, L., \& Theuns, T., 2008, MNRAS, 390, 920

\bibitem[Pe{\~n}arrubia et al.(2008)]{2008ApJ...673..226P} Pe{\~n}arrubia, 
J., Navarro, J.~F., \& McConnachie, A.~W.\ 2008, \apj, 673, 226 

\bibitem[Spergel et al.(2003)]{2003ApJS..148..175S} Spergel, D.~N., et al.\ 
2003, \apjs, 148, 175 

\bibitem[Stadel(2001)]{2001PhDT........21S} Stadel, J.~G.\ 2001, 
Ph.D.~Thesis, University of Washington  

\bibitem[Strigari et al.(2007)]{2007ApJ...669..676S} Strigari, L.~E., 
Bullock, J.~S., Kaplinghat, M., Diemand, J., Kuhlen, M., 
\& Madau, P.\ 2007, \apj, 669, 676 

\bibitem[Strigari et al.(2008)]{2008Natur.454.1096S} Strigari, L.~E., 
Bullock, J.~S., Kaplinghat, M., Simon, J.~D., Geha, M., Willman, B., 
\& Walker, M.~G.\ 2008, \nat, 454, 1096, (S08)

\bibitem[Wechsler et  al.(2002)]{2002ApJ...568...52W} Wechsler, R.~H.,
Bullock, J.~S.,  Primack, J.~R., Kravtsov, A.~V., \&  Dekel, A.\ 2002,
ApJ, 568, 52

\bibitem[White \& Rees(1978)]{1978MNRAS.183..341W} White, S.~D.~M., \& 
Rees, M.~J.\ 1978, \mnras, 183, 341 

\bibitem[Yang et al. 2007]{Yang07}
Yang, X.H., Mo, H.J., Van den Bosch, F.C., Pasquali, A., Li, C., \& Barden, M., 2007, ApJ, 671, 153
\end{thebibliography}
\end{document}